\documentclass[10pt,twocolumn,letterpaper]{article}

\usepackage[pagenumbers]{cvpr}     

\usepackage{graphicx}
\usepackage{amsmath}
\usepackage{amssymb}
\usepackage{booktabs}

\usepackage[pagebackref,breaklinks,colorlinks]{hyperref}

\usepackage[capitalize]{cleveref}
\crefname{section}{Sec.}{Secs.}
\Crefname{section}{Section}{Sections}
\Crefname{table}{Table}{Tables}
\crefname{table}{Tab.}{Tabs.}

\usepackage{cancel}
\usepackage{esint}
\usepackage{cuted}
\usepackage{caption}
\usepackage{enumitem}
\setlist[itemize]{align=parleft,left=0pt..1em}

\def\cvprPaperID{39} 
\def\confName{ICCV}
\def\confYear{2023}

\begin{document}

\title{Sensitivity analysis of AI-based algorithms for autonomous driving on optical wavefront aberrations induced by the windshield}

\author{Dominik Werner Wolf\\
{\normalsize \textbf{Volkswagen Group}}\\
{\tt\small dominik.werner.wolf}\\
{\tt\small @volkswagen.de}
\and Markus Ulrich\\
{\normalsize \textbf{Karlsruhe Institute of Technology}}\\
{\tt\small markus.ulrich@kit.de}
\and Nikhil Kapoor\\
{\normalsize \textbf{CARIAD SE}}\\
{\tt\small nikhil.kapoor}\\
{\tt\small @cariad.technology}
}
\maketitle

\begin{abstract}
Autonomous driving perception techniques are typically based on supervised machine learning models that are trained on real-world street data. A typical training process involves capturing images with a single car model and windshield configuration. However, deploying these trained models on different car types can lead to a domain shift, which can potentially hurt the neural networks performance and violate working ADAS requirements. To address this issue, this paper investigates the domain shift problem further by evaluating the sensitivity of two perception models to different windshield configurations. This is done by evaluating the dependencies between neural network benchmark metrics and optical merit functions by applying a Fourier optics based threat model. Our results show that there is a performance gap introduced by windshields and existing optical metrics used for posing requirements might not be sufficient.
\end{abstract}

\vspace{-0.6cm}
\section{Introduction}
\label{sec:intro}
The aspiration to launch level-4-ready autonomous vehicles within this decade drives new challenges in the automotive world. In order to increase the perception performance w.r.t.\ the frontal far field, the car will be equipped with high spatial resolution cameras. Advanced Driver Assistance Systems (ADAS) cameras with telephoto lenses and high-resolution sensors provide a high pixel resolution per field angle, wherefore they are more sensitive to optical aberrations within the optical path. Since car windshields are typically curved they will act as an additional lens in the optical path. Unfortunately, the curvature and thickness characteristics of the windshield are not sufficiently controllable on small domains~\cite{Windshield_Optics}. This indicates inherent optical aberrations in the optical path of ADAS cameras and impacts the recoverable information content of camera-based ADAS systems.

From the physics point of view, the imaging process of an optical system is entirely determined by the convolution of the raw image with the Point Spread Function (PSF) because of the superposition principle, which arises from the linear nature of the Helmholtz equation (\ref{eq:Helmholtz_equation}). The PSF of an aberrated optical system can be parameterized by the wavefront error in terms of Zernike coefficients~\cite{Fourier_optics}. This paper presents a methodology to define an optical threat model based on Fourier optics to reflect the perturbations induced by windshields. This difficulty becomes important if the training dataset is taken by a camera mounted on the vehicle roof but the network inference is performed on images from a camera behind the windshield. Hence, the optical aberrations can induce a significant dataset domain shift and might affect the model performance. This paper focuses on two primary research questions. First of all, how sensitive is a neural network to optical perturbations and are those sensitivities reflected sufficiently by optical merit functions, like the refractive power of the windshield for example. In order to tackle this question, we utilize a common metric in explainable AI, namely the Shapley values~\cite{Shapley}, which quantify the contribution or impact of a particular feature on the merit function of interest. The analysis of different windshield configurations will lead to a Shapley distribution for every merit function and Zernike coefficient. In order to synchronize the development efforts regarding the optical quality of windshields in the light of neural network performance, we are aiming for an optical merit function that reflects a congruent Shapley distribution as the distribution imposed by the AI benchmark metric. Secondly, we are investigating the correlation between neural network and optical system benchmark metrics. From a quality assurance perspective, we would like to determine a bijective function between neural network and optical Key Performance Indicators (KPIs). This would allow us to derive optical system requirements for the level-4 functionality. We are addressing this issue by generating different threat model attacks on the neural network architecture by Monte-Carlo sampling from uniformly distributed Zernike coefficients of second order.

The intertwining between optical characteristics and the neural network predictive power has manifested itself as a new scientific branch denoted as deep optics~\cite{Deep_optics}. The essential idea is to trim the PSF during training by minimizing the loss function. This results in the most optically informative PSF~\cite{Informative_PSF}, which might differ from the PSF with the highest imaging fidelity. For example, if the task consists of performing a depth estimation of objects by a single 2D image, it might be beneficial to code the PSF with an artificial defocusing blur~\cite{Deep_optics}. The blurring will then affect objects differently depending on their depth position. Hence, optical aberrations can be utilized as a feature for improving the information decoding capabilities of neural networks. As a downside, this methodology requires task-specific end-to-end optimization, which is not compatible with multi-task architectures typically used in the autonomous driving industry~\cite{Multi-Task_Learning_for_Autonomous_Driving_I, Multi-Task_Learning_for_Autonomous_Driving_II, Multi-Task_Learning_for_Autonomous_Driving_III}. Commonly used multi-task architectures consist of a pre-trained backbone model, which is trained on a joint dataset and is based on unified learning across multiple tasks in the encoder step. This is sequentially followed by different adaption models or simply heads that are trained on downstream, task-specific datasets, e.g., classification, segmentation and detection~\cite{Florence_Microsoft, Multi_task_network}. This hybrid architecture increases the run-time performance by the joint encoder utilization and enhances the generalization capability by incorporating data heterogeneity, which ultimately strengthens the model's robustness in inference \cite{multi_task_robustness}. As a downside, the jointly used backbone model might induce a lack of information capacity, which would result in lower task-specific KPIs~\cite{Multi-Task_Learning_for_Autonomous_Driving_III}. If we would like to make use of the deep optics approach for multi-task networks we would need to train the heads individually. As a result, the most optically informative PSF would be task-dependent, which can not be satisfied by a single optical element. Even in the case of a single-task network, the deep optics approach is economically unfeasible in the context of car windshields because of the manufacturing process, which focuses on industrial macro parameters instead of physical micro parameters that drive optical aberrations.

The results of this paper indicate that optical aberrations of the windshield can significantly deteriorate the model's performance by a domain shift and evidence on the insufficiency of existing optical working requirements for ADAS systems was found.


\section{Scope and research motivation}
For the homologation of autonomous driving vehicles, it will be necessary to perform a holistic analysis of the entire functional chain. The sensitivity analysis of image-based deep neural networks on optical aberrations induced by the windshield is only one aspect of the entire challenge. Other impact factors might also be critical, like weather conditions, out-of distribution events or lighting conditions. Those effects are not discussed in detail in this paper, which does not imply a judgment on the relative severity. The main motivation of focusing on wavefront aberrations of the windshield in this paper is based on the question: "What makes a windshield smart and level-4 ready?".

This question can only be answered if a most informative optical metric is found, which allows for deriving component requirements for safeguarding level~4 functionalities.


\section{Optical merit functions}
\label{sec:Optical merit functions}
The foundation of Fourier optics is based on the Helmholtz equation~\cite{Fourier_optics}. An electromagnetic field wave~$\rho(\Vec{x})$ has to satisfy the wave equation, which results in the time-independent Helmholtz equation:
\noindent
\begin{align}
\begin{split}
    (\bigtriangleup + k^2) \rho(\Vec{x}) &= 0 \; \; \; , \; \; \text{with:} \; \; \; k := \dfrac{2\pi f}{c} = \dfrac{2\pi}{\lambda}\;.
\end{split}
\label{eq:Helmholtz_equation}
\end{align}
\noindent
A unit amplitude spherical wave satisfies the Helmholtz equation (\ref{eq:Helmholtz_equation}) and is commonly known as the free-space Green’s function~\cite{Fourier_optics}. Generally, Green's functions are the physical version of the impulse response function in control systems engineering and are applicable to linear differential operators. If the system can be characterized by a Green's function then the system output is given by the convolution of the driving term or input signal with the Green's function. This theoretical mechanism is the causal reason for the validity of the superposition principle in optics. Therefore, the imaging process of an optical system is determined by:
\noindent
\begin{align}
\begin{split}
    \rho(\vec{x}_{o}) &= \iint_{\mathbb{R}^2} |\;h(\vec{x}_{o} - \vec{x}_{s})\;|^2 \cdot \rho(\vec{x}_{s}) \; \mathrm{dx_{s}^2}\\
    \Leftrightarrow \mathcal{F} \left[ \rho \right](\vec{k}) &= \mathcal{F} \left[ |\;h\;|^2 \right](\vec{k}) \cdot \mathcal{F} \left[ \rho \right](\vec{k})\;.
\end{split}
\label{eq:superposition_integral_incoherent}
\end{align}
\noindent
The Green's function of an optical system~$|h(\vec{x}_{o})|^2$ is commonly denoted as the PSF. It describes the image of an infinitesimal light pulse given by a Dirac delta distribution. Since we are considering an imaging system under incoherent light incidence, only the squared magnitude of the electrical field or the intensity of the light pulse matters. Essentially, with incoherent light there are no interference effects. If the Fresnel approximation is valid~\cite{Fourier_optics}, then the PSF~$|h(\vec{x}_{o})|^2$ of an optical system is determined by:
\noindent
\begin{align}
\begin{split}
    h(\vec{x}_{o}) &\approx \iint_{\mathbb{R}^2} P\left(\lambda d_{z} \vec{k}_{\Tilde{a}}\right) \cdot \mathrm{e}^{-2 \pi \mathrm{i} \vec{x}_{o} \cdot \vec{k}_{\Tilde{a}}} \; \mathrm{dk_{\Tilde{a}}^2}\\
    \Leftrightarrow h(\vec{x}_{o}) &= \mathcal{F} \left[ P\left( \lambda d_{z} \vec{k}_{\Tilde{a}} \right) \right] \; \; , \; \text{with:} \; \; \vec{k}_{\Tilde{a}} := \dfrac{\vec{x}_{a}}{\lambda \cdot d_{z}}\;.
\end{split}
\label{eq:impulse_response_final}
\end{align}
\noindent
Here,~$P(\Vec{x}_{a})$ denotes the aperture function of the optical system. In the case of an ADAS camera, the aperture stop of the objective lens is considered. Furthermore,~$d_{z}$ quantifies the distance between the observation plane at~$z_{o}$ and the position of the aperture stop at~$z_{a}$. If there are no inherent optical aberrations, then the system is diffraction limited and the incoherent impulse response function (PSF) of a one-dimensional rectangular aperture is given by the squared sinc function~\cite{Fourier_optics}. Unfortunately, optical systems in the automotive industry are not diffraction limited especially if the windshield is included. Therefore, the concept of the aperture function~$P(\Vec{x}_{a})$ has to be extended to the generalized aperture function\footnote{Generally, we adopt the Feynman slash notation if optical field quantities are assumed to be non-diffraction limited.}~${\cancel{P}}(\Vec{x}_{a})$, given by:
\noindent
\begin{equation}
    \cancel{P}(\vec{x}_{a}) := P(\vec{x}_{a}) \cdot \exp{\left[ \dfrac{2\pi\mathrm{i}}{\lambda} \cdot W(\vec{x}_{a}) \right]}\;.
\label{eq:generalized_pupil_function}
\end{equation}
\noindent
Here,~$W(\vec{x}_{a})$ denotes the wavefront aberration map on the aperture surface. Physically, the wavefront aberration map is given by the optical path difference between the expected wavefront and the observed wavefront. In the case of a windshield, the expected wavefront is given by a plane wave. Even in the case of non-diffraction limited systems, the superposition principle is applicable since the differential operator remains linear, wherefore aberrated optical systems are characterized by:
\noindent
\begin{align}
\begin{split}
    \cancel{h}(\vec{x}_{o}) &\overset{(\ref{eq:impulse_response_final})}{=} \mathcal{F} \left[ \cancel{P}\left( \lambda d_{z} \vec{k}_{\Tilde{a}} \right) \right] \\
    \Rightarrow \mathcal{F} \left[ \cancel{\rho} \right](\vec{k}) &\overset{(\ref{eq:superposition_integral_incoherent})}{=} \mathcal{F} \left[ |\cancel{h}|^2 \right](\vec{k}) \cdot \mathcal{F} \left[ \rho \right](\vec{k})\;.
\end{split}
\label{eq:superposition_integral_incoherent_aberrated}
\end{align}
\noindent
So far, the influence of optical aberrations on the imaging process has been discussed in detail and the governing physical equations were presented. For quality assurance purposes in the light of reliable autonomous driving, this physical process has to be mapped to measurable physical quantities on which quality requirements can be imposed. The following subsections elaborate on different optical metrics, which serve this objective.

\subsection{Refractive power}
Historically, refractive power measurements have been utilized as the primary quality criterion for windshields~\cite{Automotive_Glazing}. The refractive power~$D_{x_{i}}$ quantifies the curvature of the wavefront aberration map~$W(\vec{x}_{a})$ along the axis of interest~$x_{i}$ if a plane wave is expected, as it is the case for windshields~\cite{ITSC}. Consequently,~$D_{x_{i}}$ is given by:
\noindent
\begin{equation}
    D_{x_{i}}(\vec{x}_{a}) = \dfrac{\partial^2}{\partial x_{i}^2} W(\vec{x}_{a})\;.
\label{eq:refractive_power}
\end{equation}
\noindent
Current optical requirements in terms of the refractive power are typically expressed as the maximum absolute value over both transversal axes.

\subsection{Modulation Transfer Function (MTF)}
Generally, the Green's function entirely determines the behaviour of a Linear and Time-Invariant (LTI) system~\cite{Control_Systems}. As a consequence, it is insightful to further analyse the PSF. The MTF is defined as the real part of the Fourier transform of the PSF normalized to one at~$\Vec{k} = \vec{0}$, as stated by:
\noindent
\begin{equation}
     \cancel{\mathrm{MTF}}(\vec{k}) = \dfrac{\Bigr|\;\mathcal{F} \left[ |\cancel{h}|^2 \right](\vec{k})\;\Bigr|}{\Bigr|\;\mathcal{F} \left[ |\cancel{h}|^2 \right](\vec{k}=0)\;\Bigr|}\;.
\label{eq:MTF}
\end{equation}
\noindent
Hence, if the real-valued intensity PSF is considered as a light distribution in the observer plane then the MTF corresponds to the characteristic function in statistics~\cite{Green_as_PDF}, which determines the moments of the light distribution, e.g., gray values centroid, intensity variance etc. Tier-1 ADAS suppliers are recently defining functional requirements in terms of the MTF at a spatial frequency of half-Nyquist.

\subsection{Strehl Ratio (SR)}
Instead of specifying only a single spatial frequency requirement for the MTF it might be advisable to consider the entire spectrum. In order to do so, the area under the MTF curve can be evaluated. The spectral integral of the MTF in relation to the diffraction limited MTF area is defined as the Strehl ratio~\cite{Fourier_optics} and is given by:
\noindent
\begin{equation}
    \mathrm{SR}_{x_{i}} := \dfrac{\int_{\mathbb{R}} \cancel{\mathrm{MTF}}(k_{x_{i}}) \; \mathrm{dk_{x_{i}}}}{\int_{\mathbb{R}} \mathrm{MTF}(k_{x_{i}}) \; \mathrm{dk_{x_{i}}}} \overset{(\ref{eq:MTF})}{=} \dfrac{| \; \cancel{h} \; |^2}{| \; h \; |^2}\Biggr|_{\Vec{x}_{o} \overset{!}{=} \vec{0}}\;.
\label{eq:Strehl_ratio}
\end{equation}
\noindent
An equivalent definition of the Strehl ratio is given by the quotient of the aberrated PSF over the diffraction-limited PSF, evaluated at the optical axis \text{($\Vec{x}_{o} \overset{!}{=} \vec{0}$)}.

\subsection{Optical Informative Gain (OIG)}
Unfortunately, there is still a drawback in the definition of the Strehl ratio because it does not incorporate the knowledge about the shape of the PSF, which entirely characterizes the optical system. Therefore, an optical merit function would be desirable that shows a dependency on higher-order moments of the PSF as well. One possible metric that considers this constraint is introduced in this paper as the Optical Informative Gain (OIG):
\noindent
\begin{equation}
    \mathrm{OIG}_{x_{i}} := \dfrac{\int_{\mathbb{R}} \left|\cancel{\mathrm{MTF}}\right|^2 \; \mathrm{dk_{x_{i}}}}{\int_{\mathbb{R}} \left|\mathrm{MTF}\right|^2 \; \mathrm{dk_{x_{i}}}} \overset{(\ref{eq:MTF})}{=} \dfrac{\int_{\mathbb{R}} \left| \; \cancel{h} \; \right|^4 \; \mathrm{dx_{o_{i}}}}{\int_{\mathbb{R}} \left| \; h \; \right|^4 \; \mathrm{dx_{o_{i}}}}\;.
\label{eq:Optical_informative_gain}
\end{equation}
\noindent
Equation (\ref{eq:Optical_informative_gain}) takes advantage of the Plancherel theorem~\cite{Plancherel_theorem}. If the OIG is evaluated by measurement data then the domain of the MTF is restricted by the Nyquist frequency. Hence, the OIG incorporates the resolution limitation given by the image sensor and relates to the amount of photonic energy, which can be spatially discriminated in relation to the diffraction-limited case.


\section{Neural network merit functions}
Previous studies on the effect of dataset shifts~\cite{mECE_data_shift_classification} and noise corruptions~\cite{ImageNet-C} on image classification underpin the importance of optical robustness analyses for autonomous driving algorithms. The impact of dataset shifts induced by optical aberrations of the windshield on traffic sign classification has already been quantified as an accuracy drop of up to ten percent~\cite{Object_detection}. In contrast, our paper focuses on the performance and the network calibration reliability for semantic segmentation. Due to the pixel-wise class prediction, it can be hypothesized that the sensitivities for optical aberrations are amplified in relation to macro-level predictions in image classification.

\subsection{Intersection over Union (IoU)}
The governing benchmark metric for semantic segmentation is given by the Jaccard similarity coefficient or also commonly known as the Intersection over Union~(IoU)~\cite{Survey_paper}. In this paper we will make use of multi-class segmentation datasets, wherefore the mean of the~$\mathrm{IoU}$ is computed over all classes ($N_{c}$), denoted as~$\mathrm{mIoU}$.

\subsection{Expected calibration Error (ECE)}
Standard neural networks typically yield non-calibrated predictions, which can be transformed into calibrated confidence scores using post-hoc calibration methods~\cite{PTS}. Nevertheless, modern neural networks tend to yield systematically overconfident predictions~\cite{Overconfident}. A metric that assesses the calibration quality of neural networks is given by the Expected~calibration~Error~(ECE)~\cite{ECE}. For non-binary datasets, the metric is generalized as the mean over all classes (mECE).

\subsection{Shapley values}
Deep convolutional neural networks are inherently highly non-linear, wherefore it is generally difficult to assess the global sensitivity of the model predictions on single input features. One way to tackle this problem is by considering the outcome of the model with and without a particular feature. If all input feature subsets~$\mathcal{S}$ are considered regarding the marginal contribution of feature~$i$ to the sub-coalition performance then the correlations between different features are inherently incorporated. Averaging the weighted marginal contribution of feature~$i$ over all possible input feature coalitions of different cardinality results in a sensitivity metric, which fulfills all fairness properties in game theory namely the efficiency-, symmetry-, linearity- and the null player condition~\cite{Shapley_fairness}. This sensitivity metric was initially introduced by Shapley~\cite{Shapley} in the field of economics and has been widely adopted in the explainable Artificial Intelligence (AI) world since an approximative evaluation method was found by Lundberg~$\&$~Lee~\cite{Shapley_NP_hard}. In general, the Shapley value~$\varphi$ for feature~$i$ and objective function~$\hat{\Xi}$ is determined under a particular feature set~$\mathcal{M}_{f}$. Hence, the Shapley value is a local explanation method~\cite{Shapley_interpretability}, which describes the feature effect by quantifying the direction and magnitude of the local gradient in the feature space. As a consequence, if the entire feature space is sampled equidistantly the Shapley values will generate a distribution. The shape of the Shapley distribution for feature~$i$ in contrast to feature~$j$ might indicate differences in the feature importance for the neural network inference. The Shapley values:
\noindent
\begin{equation}
    \varphi_{i}(\hat{\Xi}) := \sum\limits_{\mathcal{S} \subseteq \mathcal{M}_{f}\backslash\left\{i\right\}} \resizebox{0.51\linewidth}{!}{$
    \begin{pmatrix}
        |\mathcal{M}_{f}|-1\\
        |\mathcal{S}|
    \end{pmatrix}^{-1} \dfrac{\left[ \hat{\Xi}\left(\mathcal{S} \cup \left\{ i \right\}\right) - \hat{\Xi}\left(\mathcal{S}\right) \right]}{|\mathcal{M}_{f}|}$}\;,
\label{eq:Shapley_value}
\end{equation}
\noindent
are determined by weighting the individual coalition merit with the inverse of the binomial coefficient, which quantifies the number of sub-coalitions with cardinality~$|\mathcal{S}|$.


\section{Experimental setup}
In order to examine the sensitivities and dependencies of semantic segmentation predictions on optical wavefront aberrations induced by the windshield, a proper testing environment has to be established. First of all, we will elaborate on the physical imaging model used in this paper, which utilizes Fourier optical principles to translate the wavefront aberrations induced by the windshield into image degradations. Secondly, the network architectures are introduced for conducting the evaluation experiments.

\subsection{Optical threat model}
The optical threat model, which simulates the optical aberrations of the windshield, is based on Fourier optics~\cite{Fourier_optics}. Inspired by the work of Chang et al.~\cite{Deep_optics}, we extend the proposed optical threat model to 4k ADAS cameras with telephoto objective lenses. Generally, the optical threat model assumes that the wavefront aberration map~$W(\vec{x}_{a})$ is known in advance either by measurement or optical simulation. The wavefront aberrations are parameterized by a set of Zernike coefficients~$\left\{ \omega_{n} \right\}$, which decompose the wavefront aberration map~$W$ on the unit circle:
\noindent
\begin{equation}
    W(\rho,\;\phi) = \sum \limits_{n=0}^{\infty} \omega_{n} Z_{n}(\rho,\;\phi) \;\;\;\text{,}\;\;\; \omega_{n} := \left<W,\;Z_{n}\right>\;.
\label{eq:Zernike_decomposition}
\end{equation}
\noindent
In this paper, the aperture stop of the objective lens is circular, wherefore the orthonormal\footnote{The measured Zernike coefficients might need to be renormalized according to ISO24157~\cite{ISO24157}.} Zernike polynomials~$Z_{n}$ are selected as a basis\footnote{If a square aperture stop is considered, then the Legendre polynomials would be utilized as a basis for the decomposition of~$W$.} obeying the orthogonality relation $\left<Z_{n},\;Z_{m}\right> = \delta_{nm}$. Eq.~(\ref{eq:Zernike_decomposition}) is parameterized by the normalized radius~$\rho$ and the polar angle~$\phi$ of the circular aperture.

In general, Zernike polynomials of zeroth- and first-order only induce a phase modulation, which does not impact the measured intensity distribution on the image sensor \cite{ITSC}. As a result, the incoherent MTF is not affected by the Zernike coefficients~$\omega_{0}$ to~$\omega_{2}$. This is physically sound because the zeroth order term describes a longitudinal offset of the wavefront, which does not influence the image. Secondly, the first-order terms physically describe a deflection of the light beam, wherefore the image is displaced but not structurally perturbed since the wavefront curvature is not affected. As a consequence, the studies of this paper are restricted to second-order Zernike coefficients. Higher-order terms are neglected because the magnitude of the coefficients decays with increasing order, which reflects the convergence of the series expansion in Equation (\ref{eq:Zernike_decomposition}). Future studies might also investigate terms of the truncation order, e.g., coma and trefoil.

With the knowledge of the wavefront aberration map of the windshield and the aperture stop of the camera under consideration, the generalized aperture function~${\cancel{P}}$ can be constructed by applying Equation (\ref{eq:generalized_pupil_function}). Based on~${\cancel{P}}$ the incoherent, non-diffraction limited PSF~$|{\cancel{h}}|^2$ is computed based on Equation (\ref{eq:superposition_integral_incoherent_aberrated}), which entirely characterizes the optical system. The perturbed image~${\cancel{\rho}}$ is finally given by convolving the clean image~$\rho$ with the perturbed PSF~$|{\cancel{h}}|^2$. From the measured wavefront aberration map and the deduced PSF, the entire ensemble of optical merit functions introduced in Section~\ref{sec:Optical merit functions} can be derived.

\begin{figure}[b]
  \centering
  \vspace{-0.3cm}
  \includegraphics[width=1.\linewidth]{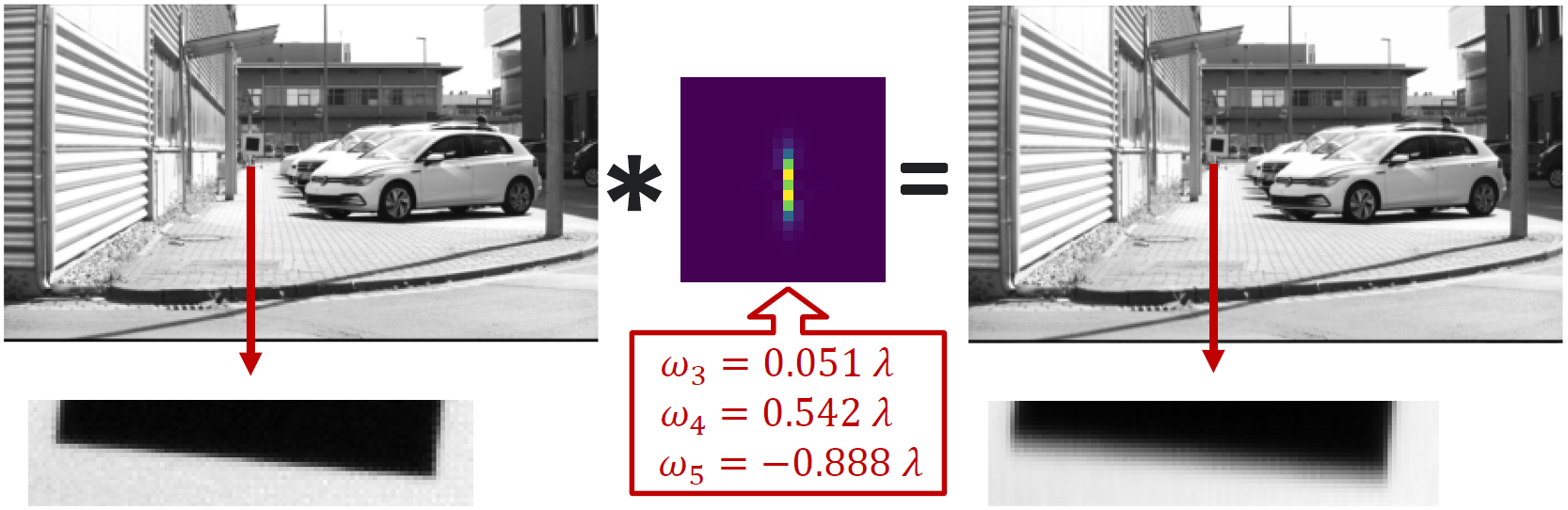}
  \caption{Toy example demonstrating the effect of the optical threat model applied to a real-world scene. The slanted edge targets are shown enlarged on the bottom.}
  \label{fig:threat_model}
  \vspace{-0.3cm}
\end{figure}

\begin{figure}[t]
  \centering
  \includegraphics[width=1.\linewidth]{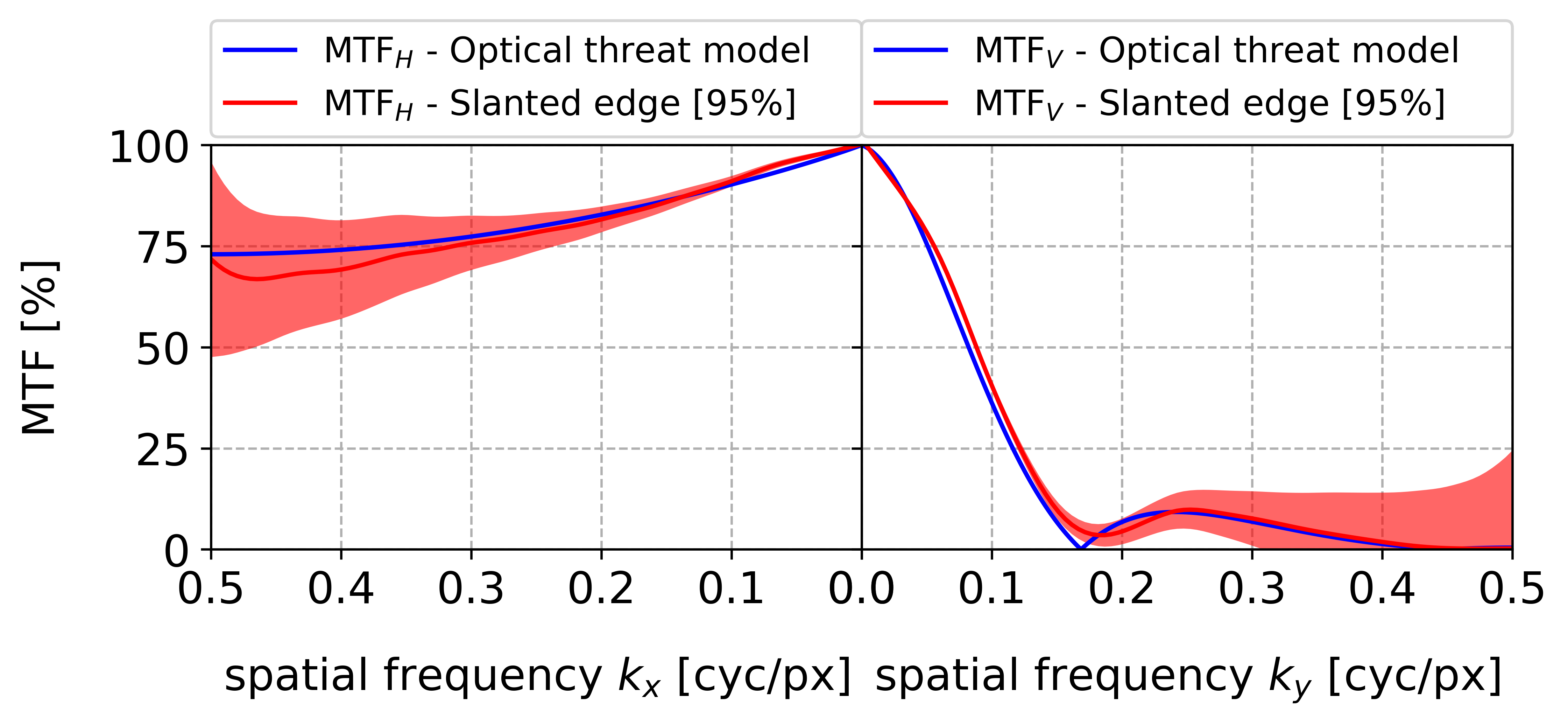}
  \vspace{-0.5cm}
  \caption{Validation of the optical threat model based on MTF measurements with the slanted edge method. The confidence bands are reflecting the Poisson noise of the image sensor.}
  \label{fig:threat_model_validation_MTF}
\end{figure}

Figure~\ref{fig:threat_model} demonstrates the effect of the optical threat model.
\noindent
\noindent
The Zernike coefficients for the parameterization of the wavefront aberration map were determined by a Shack-Hartmann wavefront measurement of a test sample windshield. The black square target within the image has been utilized for a slanted edge analysis according to ISO12233~\cite{ISO12233}. The MTF of the perturbed image is normalized by the MTF of the undistorted image to retrieve the net effect of the induced optical aberrations. The resulting MTF curves for the horizontal and vertical direction are compared to the MTF parameterized by the optical threat model in Figure~\ref{fig:threat_model_validation_MTF}. In addition, the refractive power triggered by the curvature modulation of the wavefront can be evaluated by Equation (\ref{eq:refractive_power}), which has been benchmarked by a reference refractive power measurement using the Moiré pattern technique~\cite{Metrologia}. In conclusion, the measurement results for the physical test sample are sufficiently reflected by the optical threat model, which underpins the validity of the implemented Fourier optics approach.
\noindent
\noindent

\vspace{-0.1cm}
\subsection{Architecture of the evaluation networks}
In this paper, we make use of a publically available deep convolutional neural network trained on the KITTI dataset~\cite{KITTI} from TensorFlow Hub and we study a state-of-the-art multi-task network from CARIAD.

\vspace{-0.2cm}
\subsubsection{High-Resolution Network (HRNet)}
\vspace{-0.1cm}
The High-Resolution Network (\text{HRNet}) architecture has been invented by Microsoft~\cite{HRNet_Microsoft} and the TensorFlow Hub model was adapted and trained by Google~\cite{HRNet_Google_I, HRNet_Google_II}. The selection of the \text{HRNet} architecture as an evaluation model is based on the fact that future ADAS functionalities will most likely rely on 4k high-resolution cameras. In general, a model architecture can be tuned in three dimensions: depth (e.g., more layers), wideness (e.g., more channels) or finesse (e.g., higher resolution images). The standard sequential encoder-decoder architecture in deep convolutional neural networks lacks on information capacity in the condensed low-resolution feature map. Hence, the standard encoder-decoder architecture is typically extended for highly spatially sensitive applications like autonomous driving. The \text{HRNet} tackles this challenge by switching the information propagation from serial to parallel~\cite{HRNet_Microsoft}. In detail, convolutions are performed in parallel on multiple resolutions to improve the information capacity of the model architecture. Therefore, the high-resolution representation of the input information is maintained throughout the whole process. Repeated fusion steps between parallel streams of different resolutions ensure an information flow across the levels.

\vspace{-0.2cm}
\subsubsection{Multi-Task Learning (MTL) model}
\vspace{-0.1cm}
The in-house developed Multi-Task Learning (MTL) model consists of a large shared encoder with several feature extraction layers followed by five decoder heads, each for a specific task, referred to as task head. These task heads are mainly of two types: segmentation heads and object detection heads. In detail, the parallelized decoders address the following tasks:

\begin{itemize}
    \item \textbf{Semantic segmentation head:} Provides a pixel-wise classification across the image for several classes. The head's performance is quantified by the mIoU.
    \item \textbf{Blockage detection head:} Provides a binary segmentation mask that detects if a certain region of the image is blocked or not. The evaluation metric is given by the IoU.
    \item \textbf{Traffic Light Detection and Classification:} At first, 2D bounding boxes for traffic lights in the image are predicted. Subsequently, the pixels within a single 2D bounding box are segmented to either belong to the class "traffic light bulb" or "housing". Finally, the pixels belonging to the class "traffic light bulb" are used to classify the signal color of the corresponding traffic light. For quantifying the performance of this multi-step classification task, a head-specific combined metric is evaluated, which relies, i.a., on the average accuracy and the area under the precision-recall curve.
    \item \textbf{Traffic Sign Detection and Classification:} Predicts 2D bounding boxes for traffic signs within the image. Afterwards, the subimages are used to classify the corresponding traffic sign type. Similar to the traffic light classification head a combined metric is assessed for the performance of the multi-step traffic sign classification task.
    \item \textbf{Vehicle Detection:} Provides a categorized 3D bounding box across two types of vehicles: large vehicles (e.g., trucks, buses, etc.) and passenger cars. The head's performance is evaluated by the average precision metric.
\end{itemize}
For a consolidated evaluation, we first determine the task-specific metrics (i.e., average precision for object detection and mIoU for semantic segmentation). However, to convey a holistic model performance, the head-specific loss functions are integrated using weighted averaging after normalization culminating into an overall combined multi-task loss, ranging from 0 (worst performance) to 1 (perfect performance). This aggregated score reflects the model's collective efficiency across all tasks. In order to prevent a single task from being dominant in the learning process, the individual, task-specific head losses can be integrated by an uncertainty-based weighting scheme to obtain a more robust combined metric \cite{weighted_losses}.

\subsection{Evaluation datasets}
Typically, datasets for autonomous driving are taken with cameras mounted behind the windshield. As a consequence, the images are inherently perturbed by optical aberrations, which leads to an unknown dataset domain shift that makes it impossible to quantitatively assess the impact of different windshield configurations without prior knowledge about the inherent aberrations. Hence, it is eminently beneficial that the \text{HRNet} was trained on the KITTI dataset, where the camera had been mounted on the car roof \cite{KITTI}. The evaluation images from the KITTI dataset are characterized by a resolution of \text{$370 \times 1224$\;px}.

The MTL model is trained on a joint dataset, i.e., each head is trained on a task-specific dataset with corresponding labels. The multi-task dataset from CARIAD features images of the dimensions \text{$1024 \times 2048$\;px}.


\vspace{-0.1cm}
\section{Evaluation results}
\vspace{-0.1cm}
The results obtained from employing the optical threat model on two distinct neural network architectures are summarized in this section. In general, the results for the \text{HRNet} and the MTL model are primarily coherent, e.g., the dependency of the model performance on the optical merit functions introduced in Section~\ref{sec:Optical merit functions} and the network performance sensitivity on different Zernike coefficients.

\vspace{-0.1cm}
\subsection{Sensitivity analysis}
\begin{figure}[b]
   \centering
   \vspace{-0.5cm}
   \includegraphics[width=0.8\linewidth]{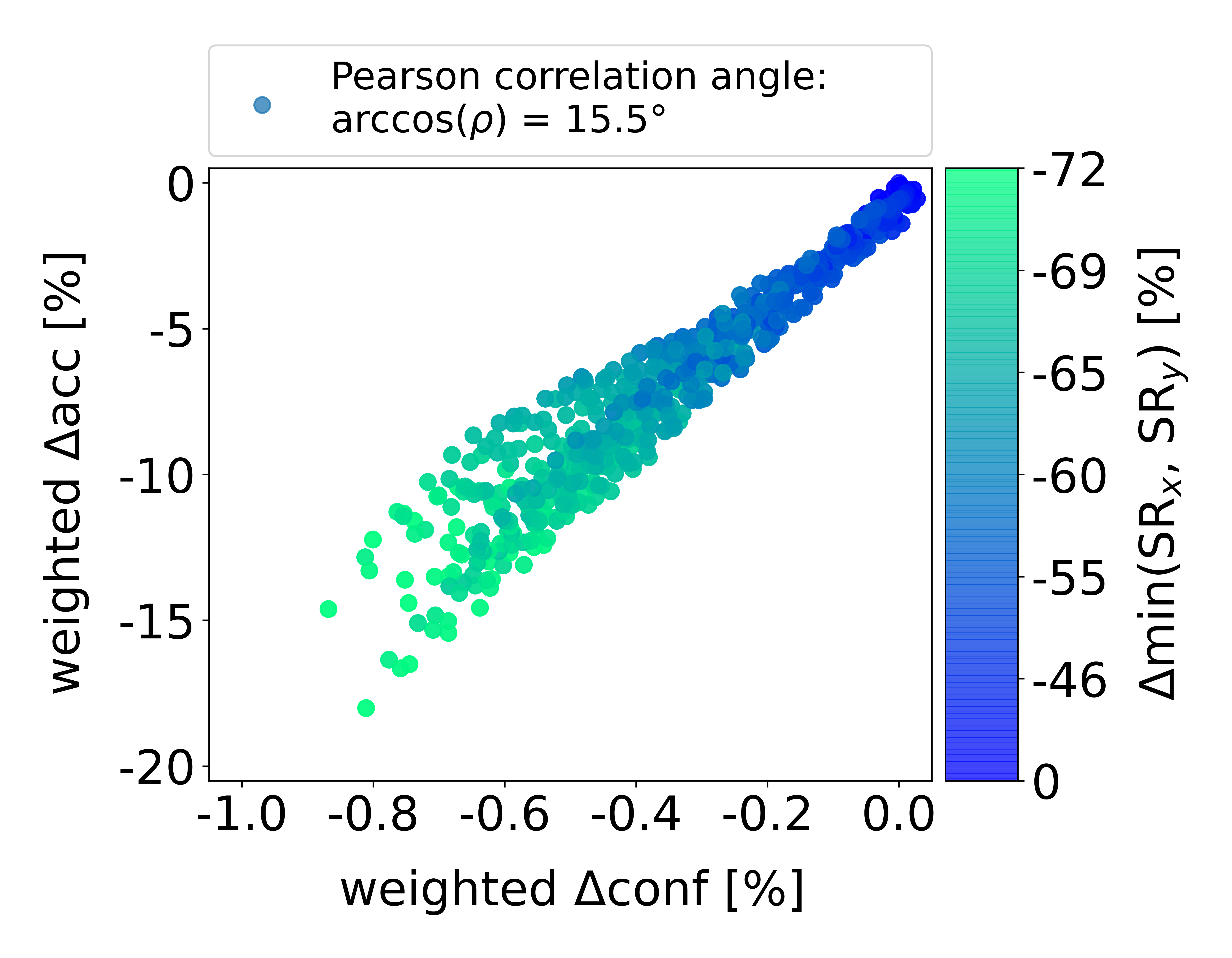}
   \vspace{-0.3cm}
   \caption{Pearson correlation between the weighted confidences and the weighted accuracies (bin cardinality weighting scheme).}
   \label{fig:pearson_correlation_conf_vs_acc}
\end{figure}

\vspace{-0.1cm}
The Shapley studies envisioned by Figure~\ref{fig:Sensitivity_analysis} on different optical merit functions and neural network benchmark measures indicate a non-linear mapping of the sensitivities between the AI world and the optical world. For comparability reasons, the Shapley values have been normalized to the effect of~$\omega_{4}$, which physically represents defocus. The behavior of the mIoU and the mECE with increasing perturbation magnitude is physically sound and predicted but the symmetry is remarkable. The mirror symmetry w.r.t.\ the abscissa originates from the observation that the mECE is dominated by the accuracy degradation, as illustrated by Figure~\ref{fig:pearson_correlation_conf_vs_acc}. In addition, the refractive power shows no sensitivity regarding~$\omega_{3}$ as expected, which reflects the fact that the merit function has been explicitly restricted to the~$x$- and~$y$-axis. In general, it can be concluded that~$\omega_{4}$ aberrations have the biggest impact on the performance of the studied merit functions. Furthermore, the Shapley distributions regarding the MTF, the Strehl ratio and the OIG are very similar in terms of their codomains but they reveal slightly different probability allocations, which indicates different statistical moments.
\noindent
\noindent

\subsection{Correlation analysis}
The dependencies between optical merit functions and neural network benchmark measures are directly contrasted in Figure~\ref{fig:Correlation_analysis_HRNet} for the HRNet.
\noindent
\begin{figure}[b]
    \centering
    \vspace{-0.6cm}
    \includegraphics[width=1.\linewidth]{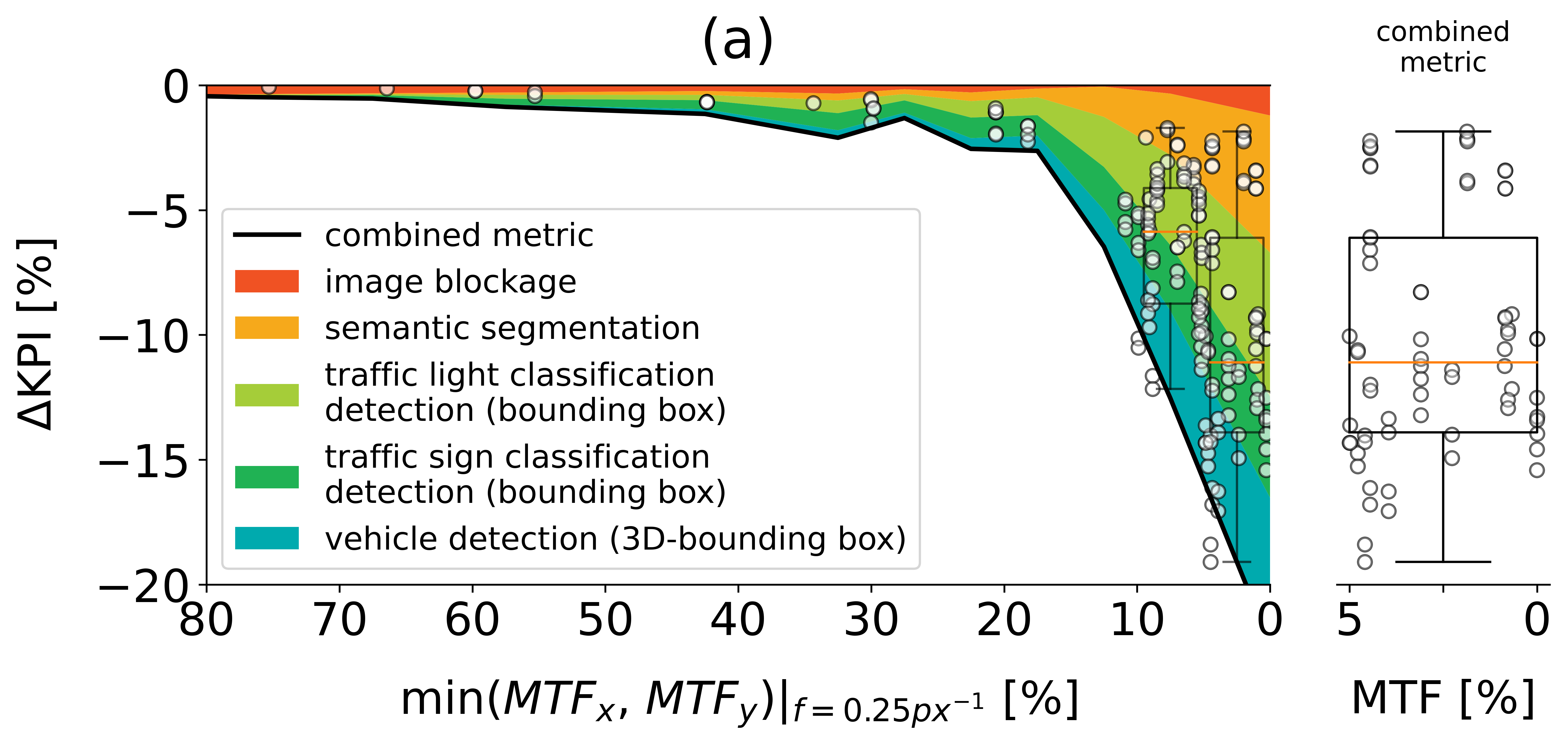}
    \includegraphics[width=1.\linewidth]{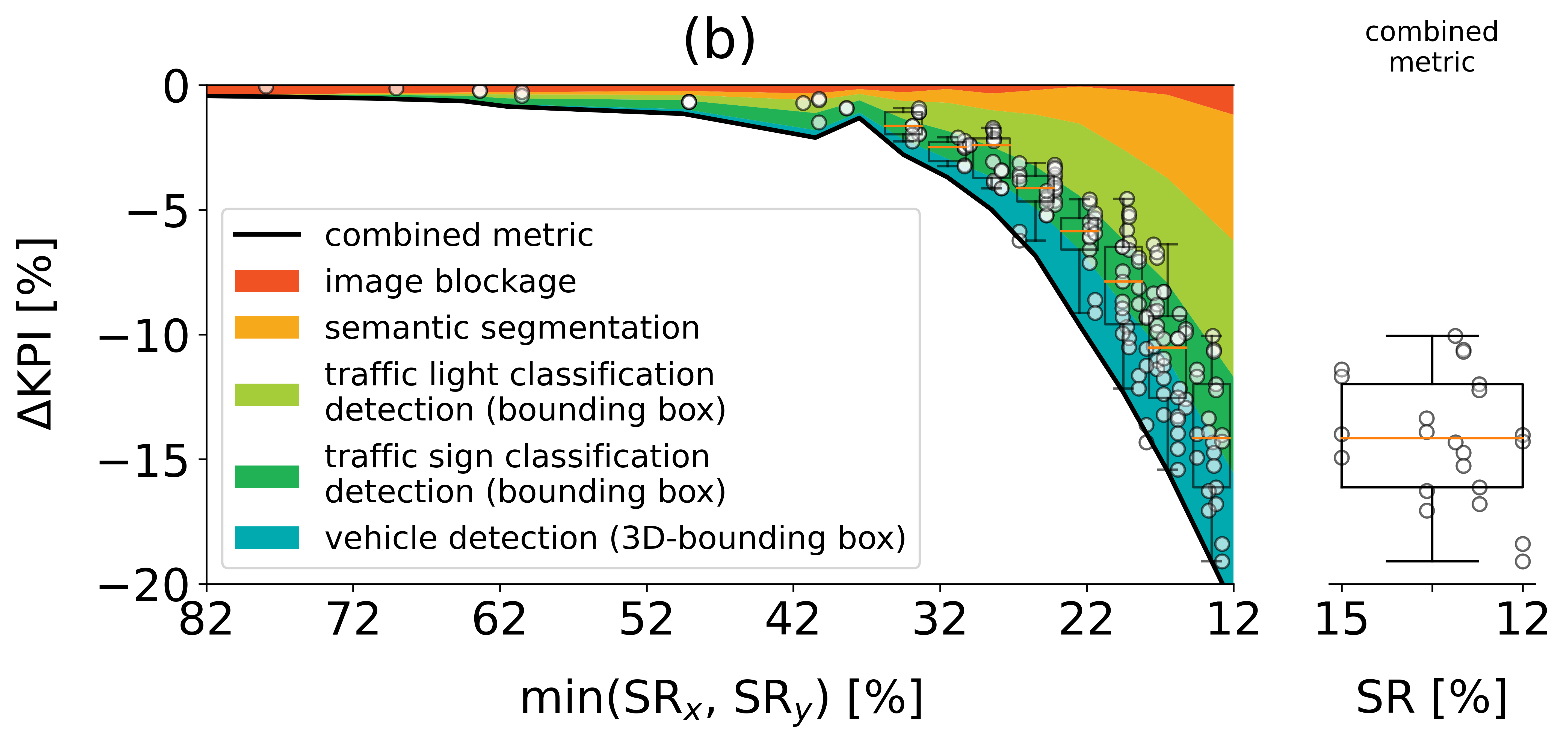}
    \includegraphics[width=1.\linewidth]{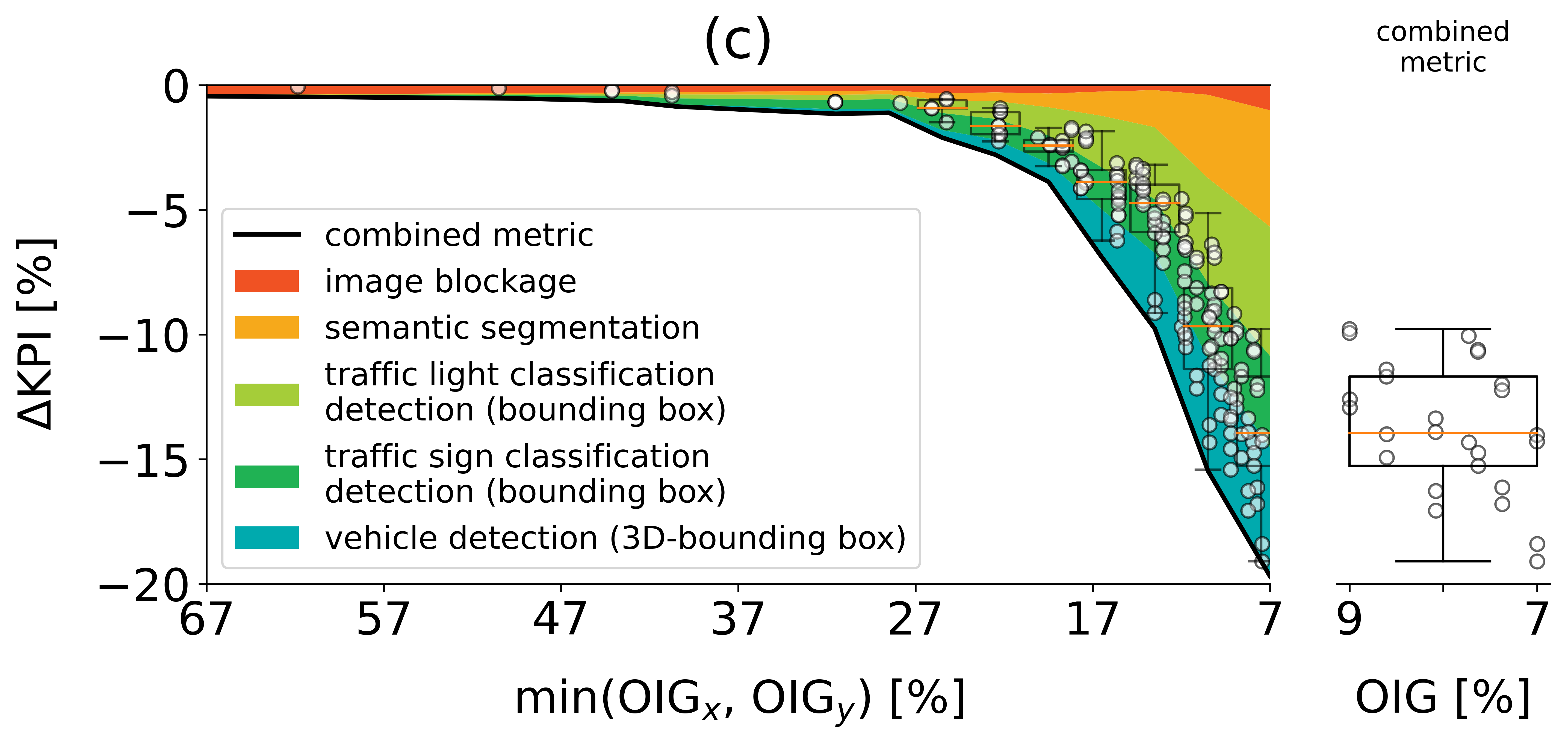}
    \vspace{-0.6cm}
    \caption{Multi-task performance versus (a) the MTF at half-Nyquist, (b) the Strehl ratio and versus (c) the OIG.}
    \vspace{-0.4cm}
\label{fig:Robusta_dependency}
\end{figure}
\noindent
It is clearly noticeable that from the refractive power and the MTF it is not possible to infer the performance of the neural network unambiguously. On the contrary, the Strehl ratio and the OIG indicate a functional relationship to the mIoU as well as the mECE within the uncertainty intervals. The uncertainty bars are given by the standard deviation of the mean of the mIoU and the mECE regarding the test image batch of size~$40$. In addition, the MTL model shows similar performance trends regarding the Fourier optical metrics as the \text{HRNet} if the envelope functions and the subdomain fluctuations in Figure~\ref{fig:Robusta_dependency} are considered. Here, each data point represents the effect of a windshield, parameterized by a set of Zernike coefficients in the range of~$\omega_{n} \in \left[-\lambda,\;\lambda\right]$, on a test image batch of size~$20$. The envelope function is obtained by binning the data and assigning the minimum value within a bin to the envelope function. This procedure can be performed for all head-specific KPIs leading to the colored stack plot in Figure~\ref{fig:Robusta_dependency} after applying the uncertainty-weighting. In order to indicate the local performance spread, additional boxplots are provided for the most relevant bins. The bin comprising the most severe optical aberrations is highlighted on the right-hand side in Figure~\ref{fig:Robusta_dependency}. It can be concluded, that the statistical mass allocations within each bin are significantly more clustered in the case of the Strehl ratio and the OIG as compared to the MTF at Nyquist half frequency. Hence, Figure~\ref{fig:Robusta_dependency} shows evidence for the superiority of the Strehl ratio and the OIG as a quality metric in contrast to the MTF at Nyquist half frequency. The results clearly indicate that the information density of the PSF is beneficial for defining an optical ADAS working requirement. For quality assurance purposes it would be required to ensure that the quality metric is bijective, which is the subject of future studies. Overall, the MTF at Nyquist half frequency seems to be insufficient as a safety quality criterion for windshields.

\subsection{Calibration robustness}
\begin{figure}[t]
  \centering
   \includegraphics[width=0.7\linewidth]{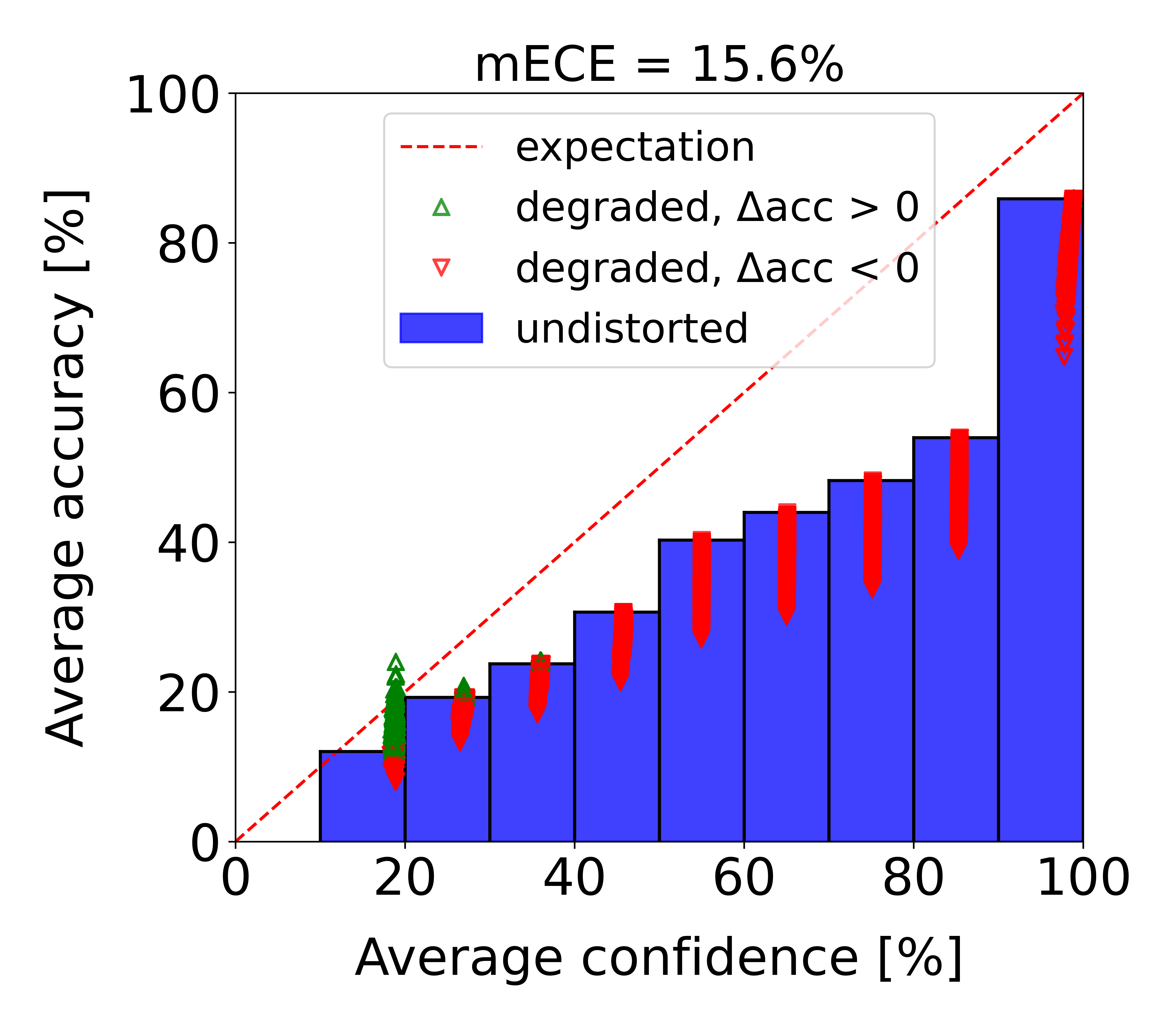}
   \vspace{-0.3cm}
   \caption{Calibration curve for the \text{HRNet} from Google.}
   \vspace{-0.1cm}
   \label{fig:calibration_curve}
\end{figure}

The effect of optical aberrations on the reliability curve of the \text{HRNet} is visualized in Figure~\ref{fig:calibration_curve}.
\noindent
\noindent
In the diffraction limited case ($\omega_{n}=0 \; \forall \; n \in \mathbb{N}_{0}$), the \text{HRNet} shows an mECE of~$15.6\%$. If optical aberrations are considered, then the average accuracy and the average confidence decrease with increasing perturbation magnitude but the reduction is non-coherent, as demonstrated by Figure~\ref{fig:pearson_correlation_conf_vs_acc}. Consequently, the neural network becomes more and more overconfident, which is underpinned by the increasing mECE in Figure~\ref{fig:Sensitivity_analysis}. In the case of low prediction confidences and low prediction accuracies, the binned network accuracy seems to slightly increase if aberrated test data is used. This behavior is counterintuitive and represents an artifact of the visualization method. Essentially, optical aberrations drive the probability flow of the confidence distribution towards lower values shifting the predictions into low-confidence bins. Since the domain is limited and quantized, the bin composition varies affecting the binned accuracy. Physically, low-confidence predictions in semantic segmentation mostly correspond to class area borders. If the image is perturbed, then the contours are getting blurred as illustrated by Figure~\ref{fig:threat_model}, which results in lower prediction confidences for these pixels.

\clearpage
\newpage


\section{Conclusion}
The results presented in this paper reflect an initial assessment of the functional relationship between neural network benchmark metrics and optical aberrations parameterized by Zernike coefficients. From our experiments, we report evidence on the superiority of the Strehl ratio and the OIG as an optical quality indicator for image-based neural network performance in contrast to present functionality requirements in terms of the refractive power and the MTF at half-Nyquist frequency. In addition, the studies demonstrate that a pure defocus is influencing the performance of a semantic segmentation algorithm more than astigmatic aberrations. Furthermore, the investigations on the \text{HRNet} from Google and the studies on the MTL model from CARIAD show similar sensitivities and functional dependencies on optical aberrations, which leads to the hypothesis, that the results presented in this paper could be network architecture independent.

Finally, it has to be emphasized that the optical threat model applied in this paper was tuned for telephoto objective lenses. As a consequence, the scalar product in the exponent of Equation (\ref{eq:impulse_response_final}) is assumed to be given by the product of the vector magnitudes. If wide-angle cameras are considered, then the optical threat model has to be adjusted by including the dependency on the field angle~$\psi$ as: ${\vec{x}_{o} \cdot \vec{k}_{\Tilde{a}} = ||\vec{x}_{o}||_{2} \cdot ||\vec{k}_{\Tilde{a}}||_{2} \cdot \cos{\psi}}$.

\begin{strip}
\vspace{-0.3cm}
\begin{center}
      \includegraphics[width=0.33\textwidth]{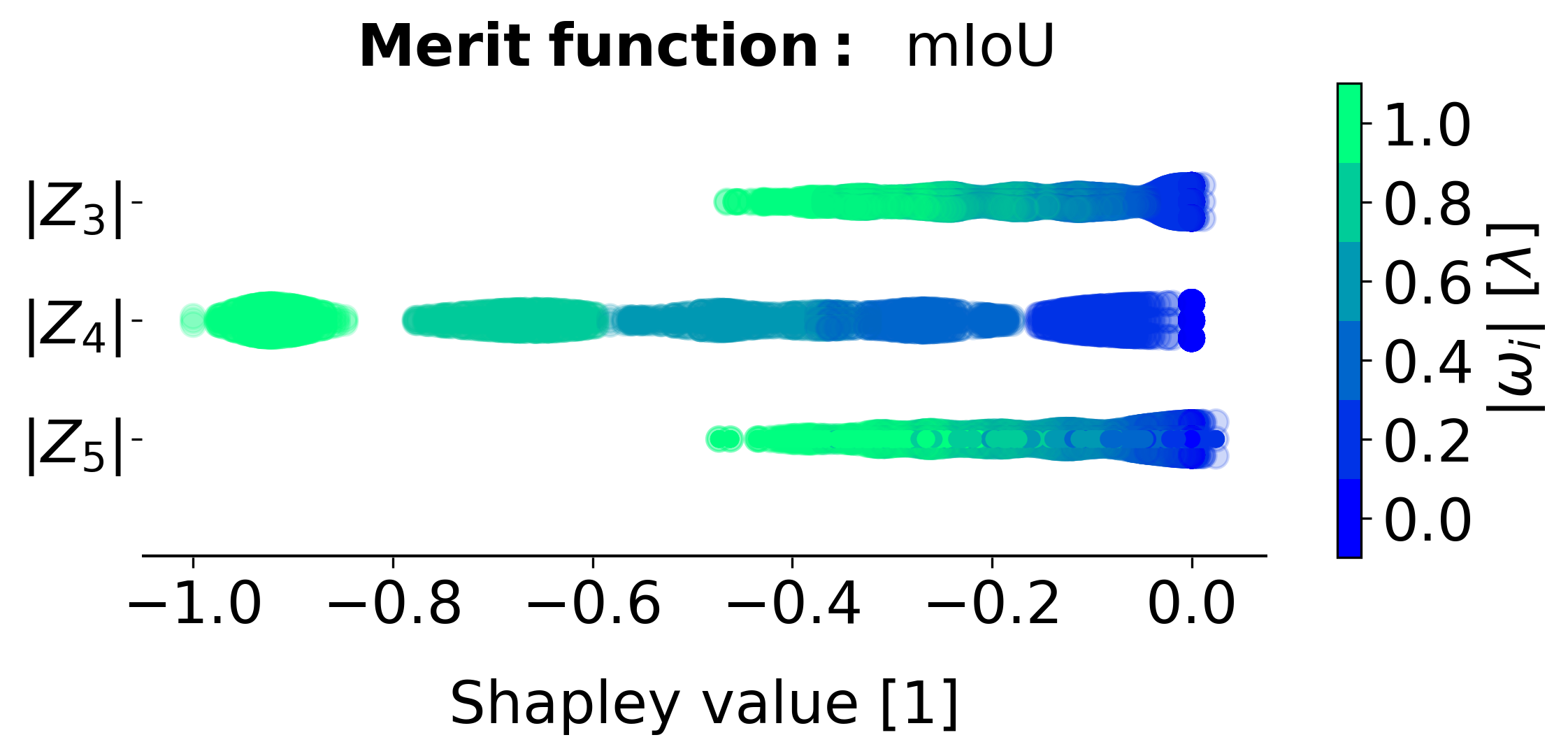}
      \hfill
      \includegraphics[width=0.33\textwidth]{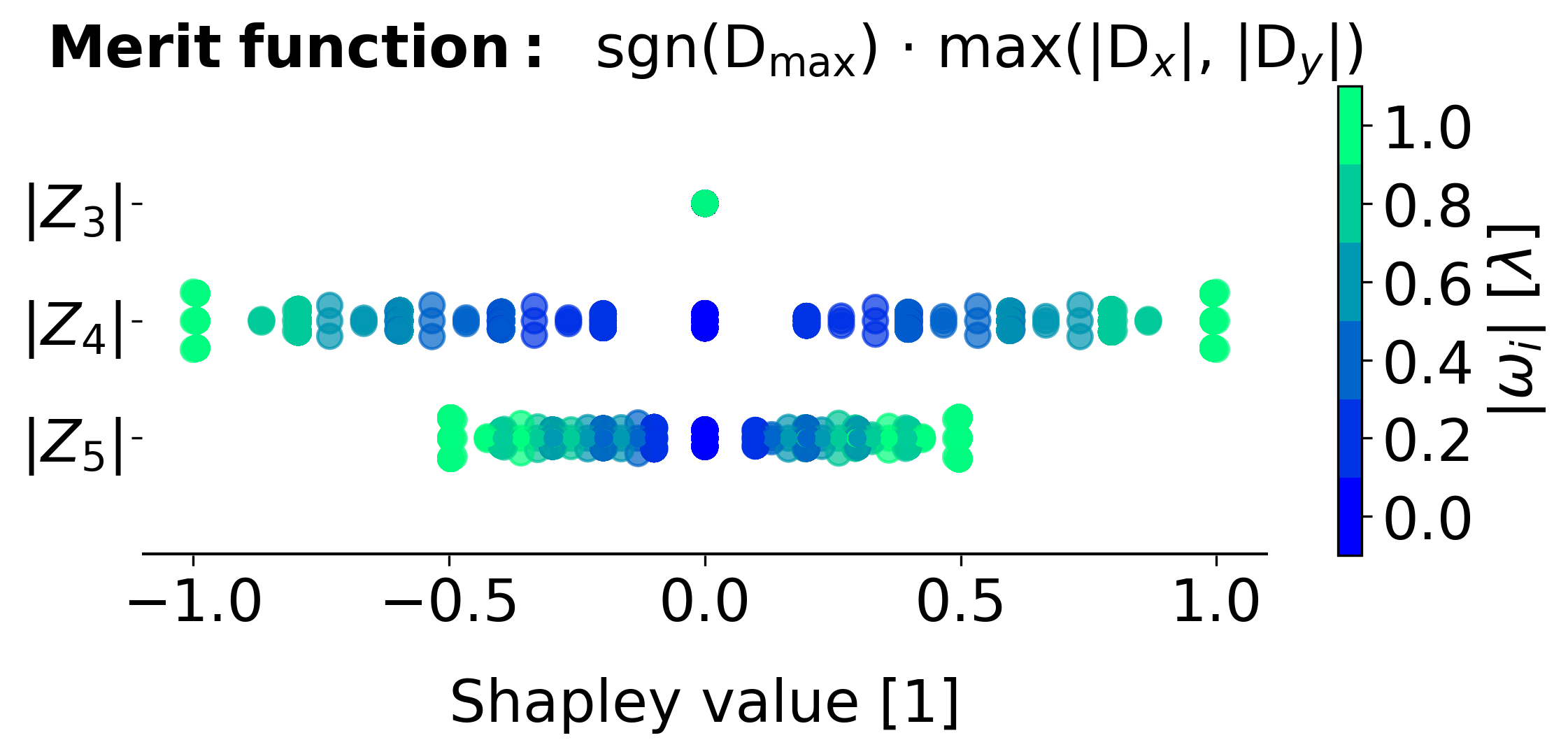}
      \hfill
      \includegraphics[width=0.33\textwidth]{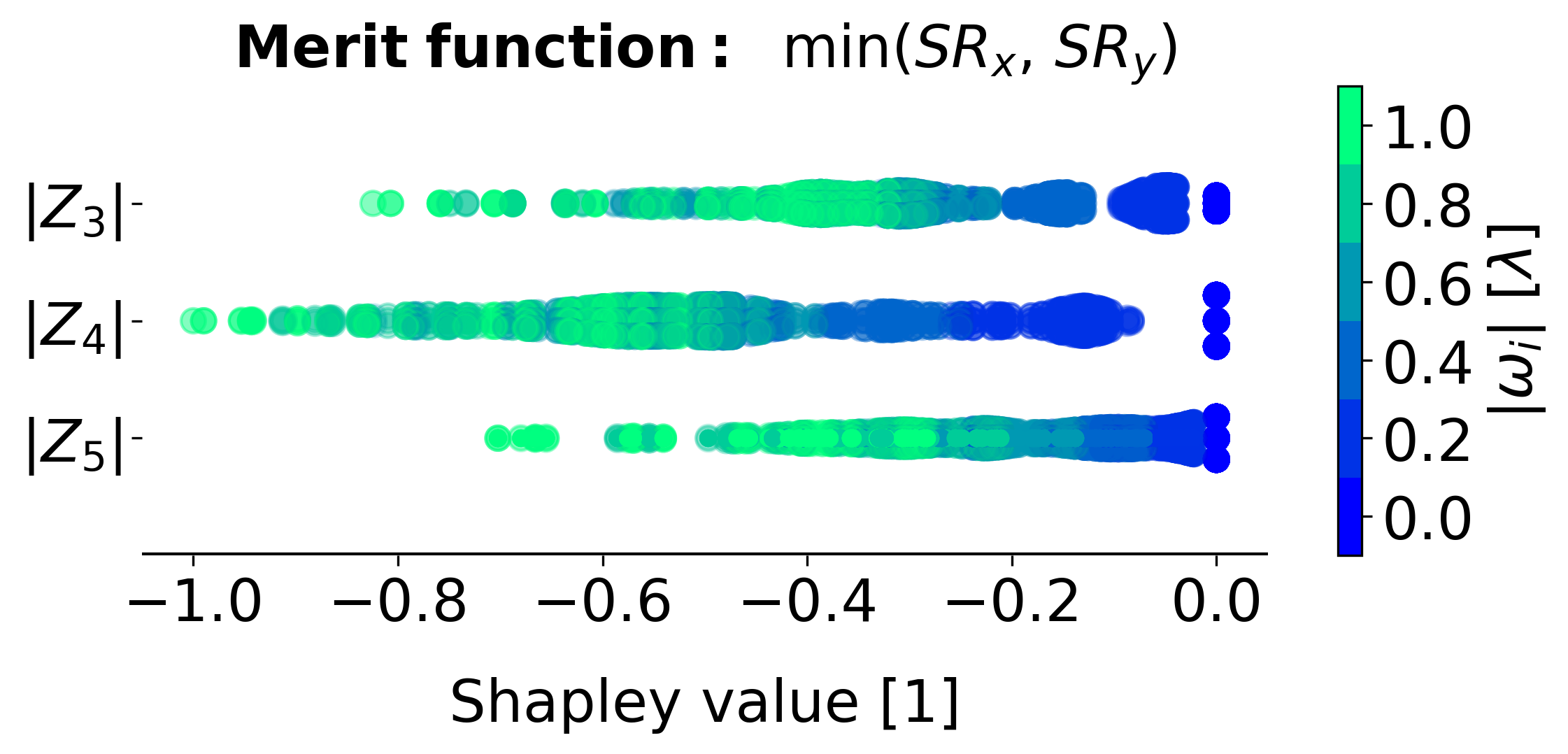}\\[5pt]
      \includegraphics[width=0.33\textwidth]{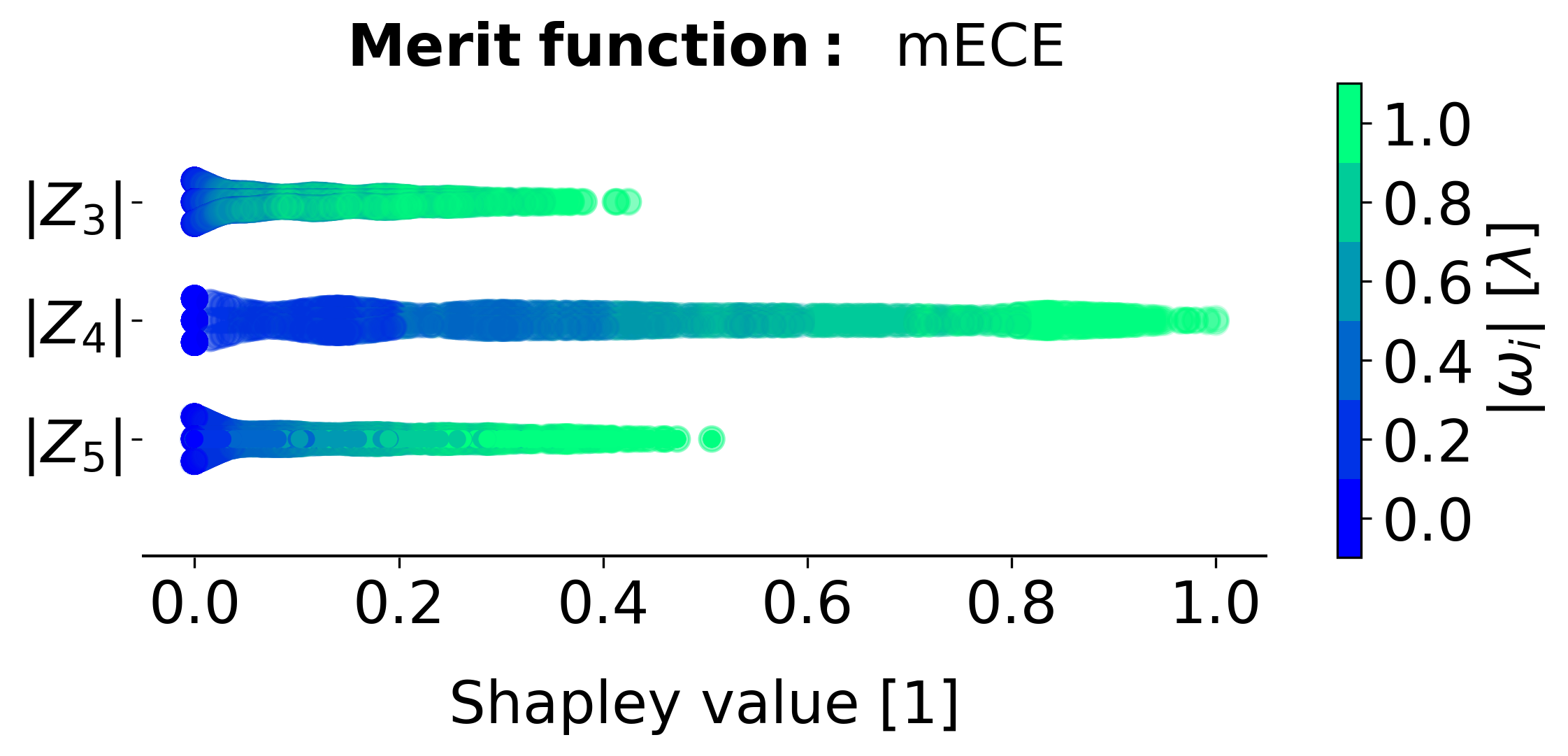}
      \hfill
      \includegraphics[width=0.33\textwidth]{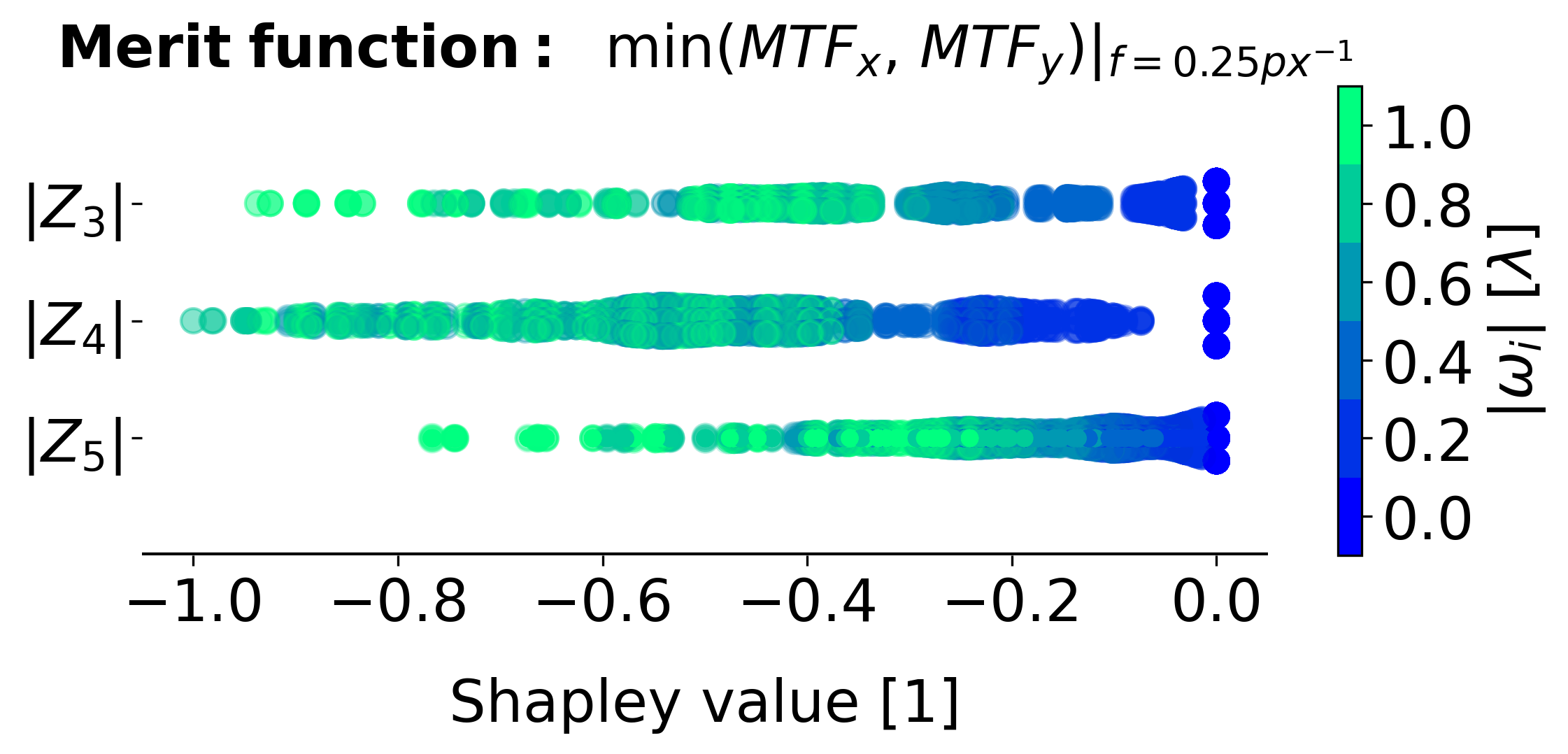}
      \hfill
      \includegraphics[width=0.33\textwidth]{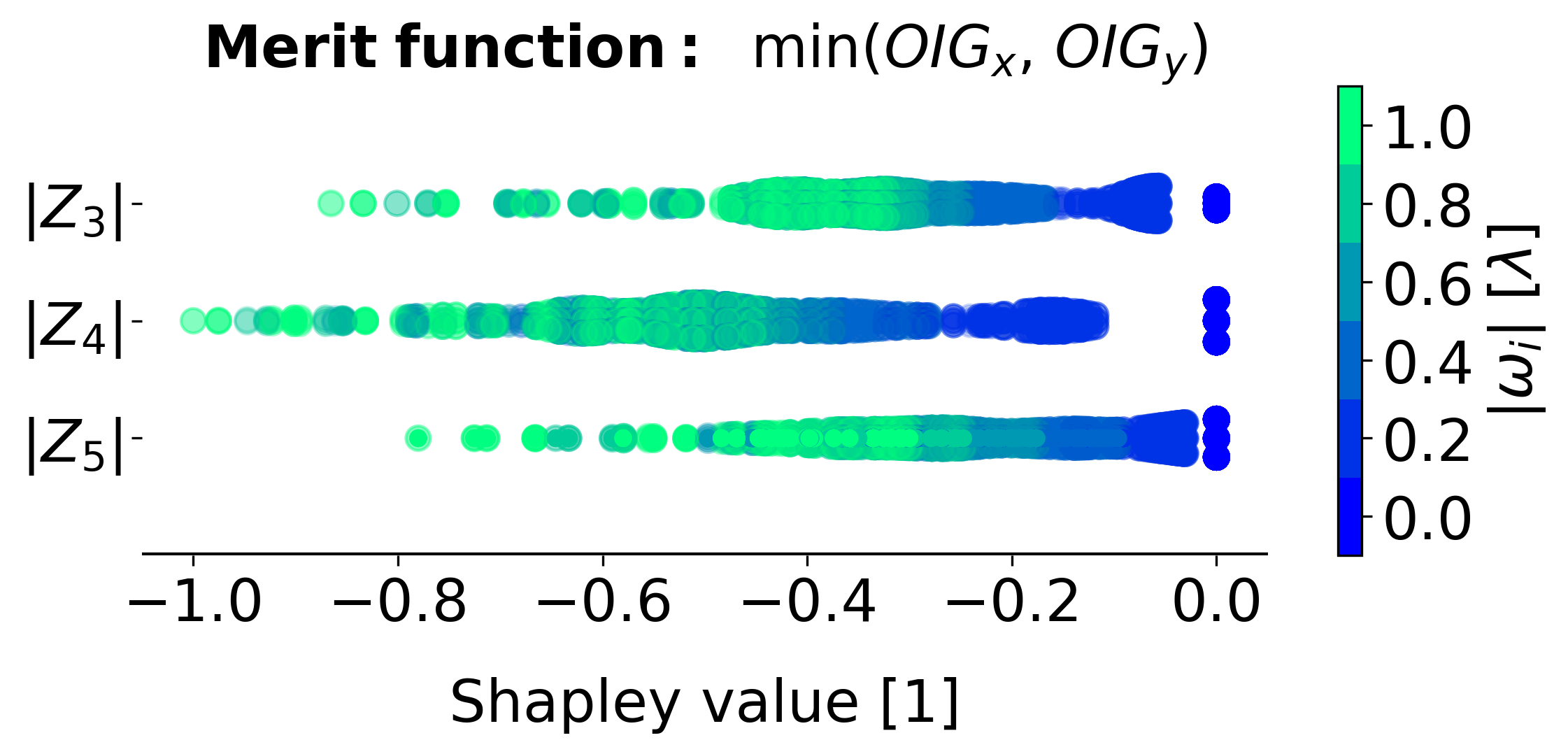}
\captionof{figure}{The sensitivities of different convolutional neural network benchmark metrics as well as the sensitivities of several optical KPIs on wavefront aberrations, parameterized by Zernike coefficients, are quantified and visualized in terms of Shapley values. The impact of an induced defocus~(Z$_{4}$) surpasses the effect of oblique-~(Z$_{3}$) and vertical astigmatism~(Z$_{5}$) for all merit functions studied in this paper.}
\label{fig:Sensitivity_analysis}
\end{center}
\begin{center}
      \hfill
      \includegraphics[width=0.41\textwidth]{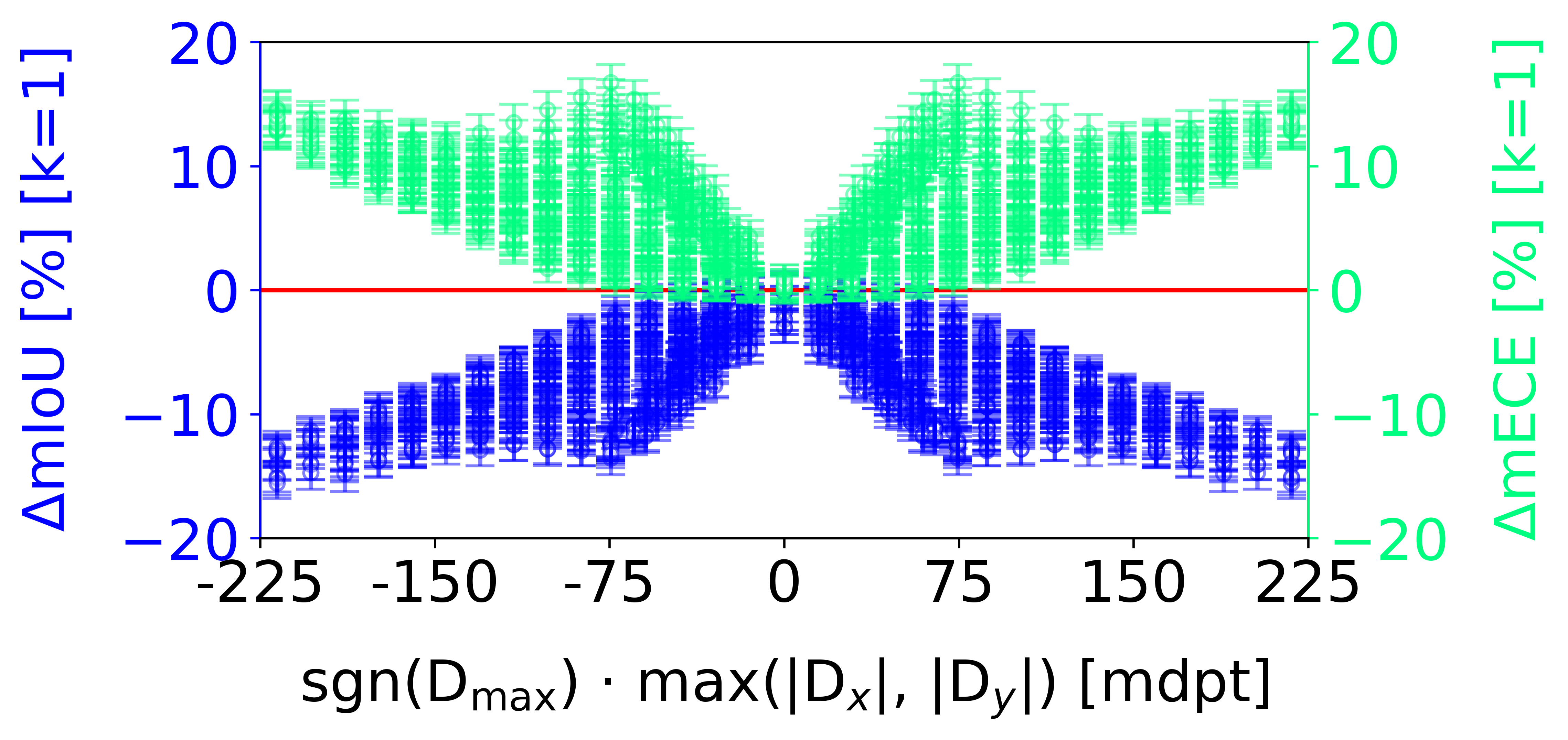}
      \hfill
      \includegraphics[width=0.41\textwidth]{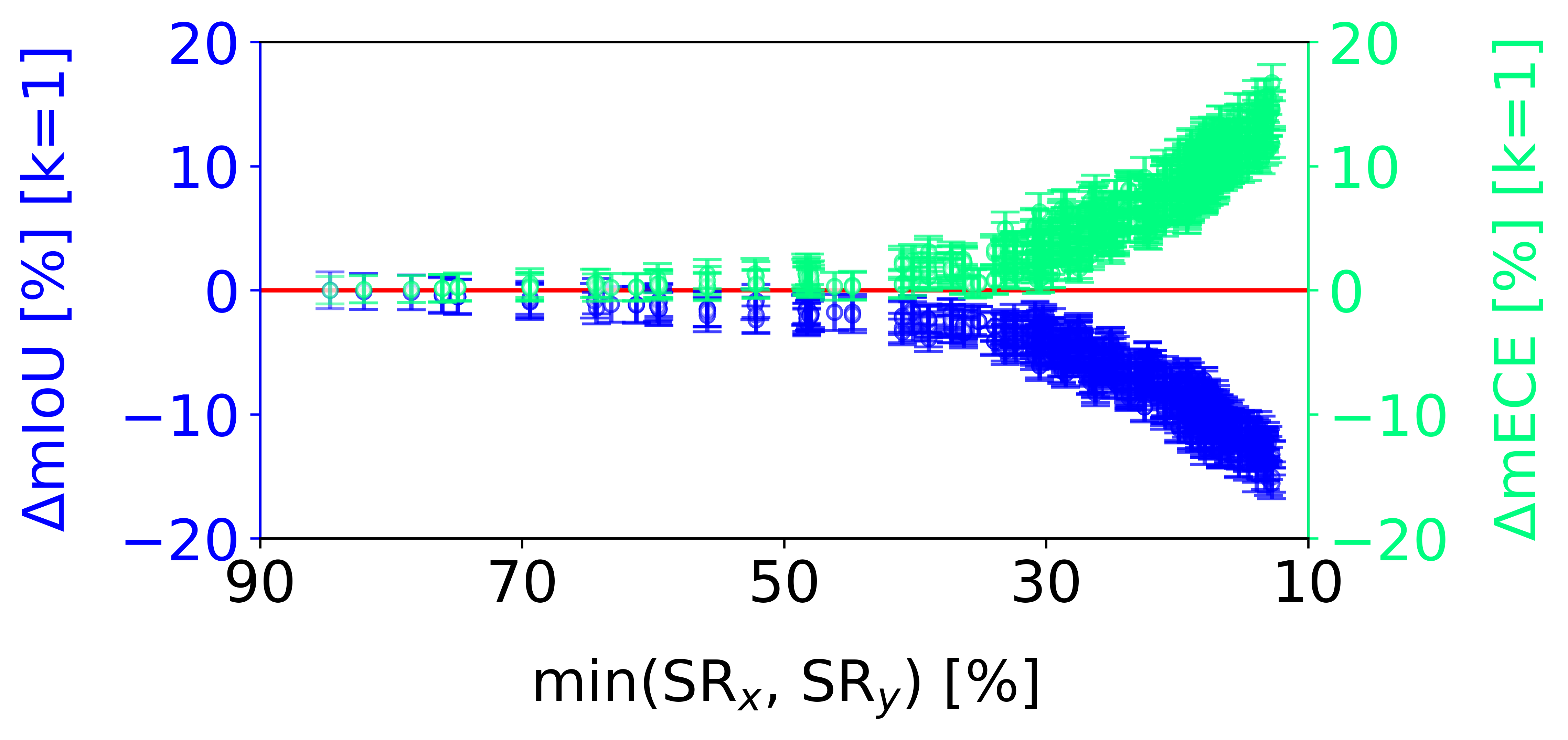}
      \hfill
      \includegraphics[width=0.\textwidth]{mIoU_vs_Strehl_ratio.png}\\[2pt]
      \hfill
      \includegraphics[width=0.41\textwidth]{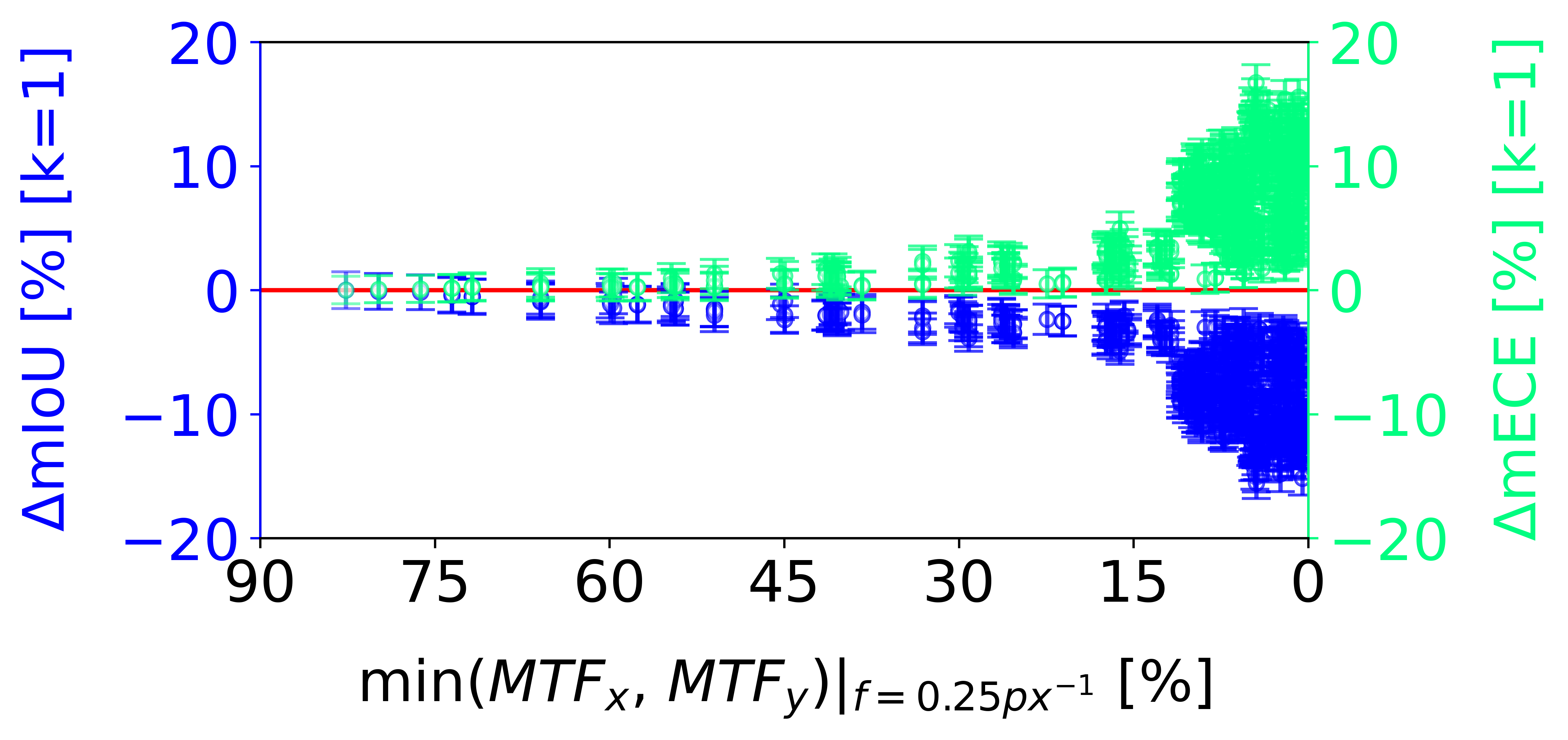}
      \hfill
      \includegraphics[width=0.41\textwidth]{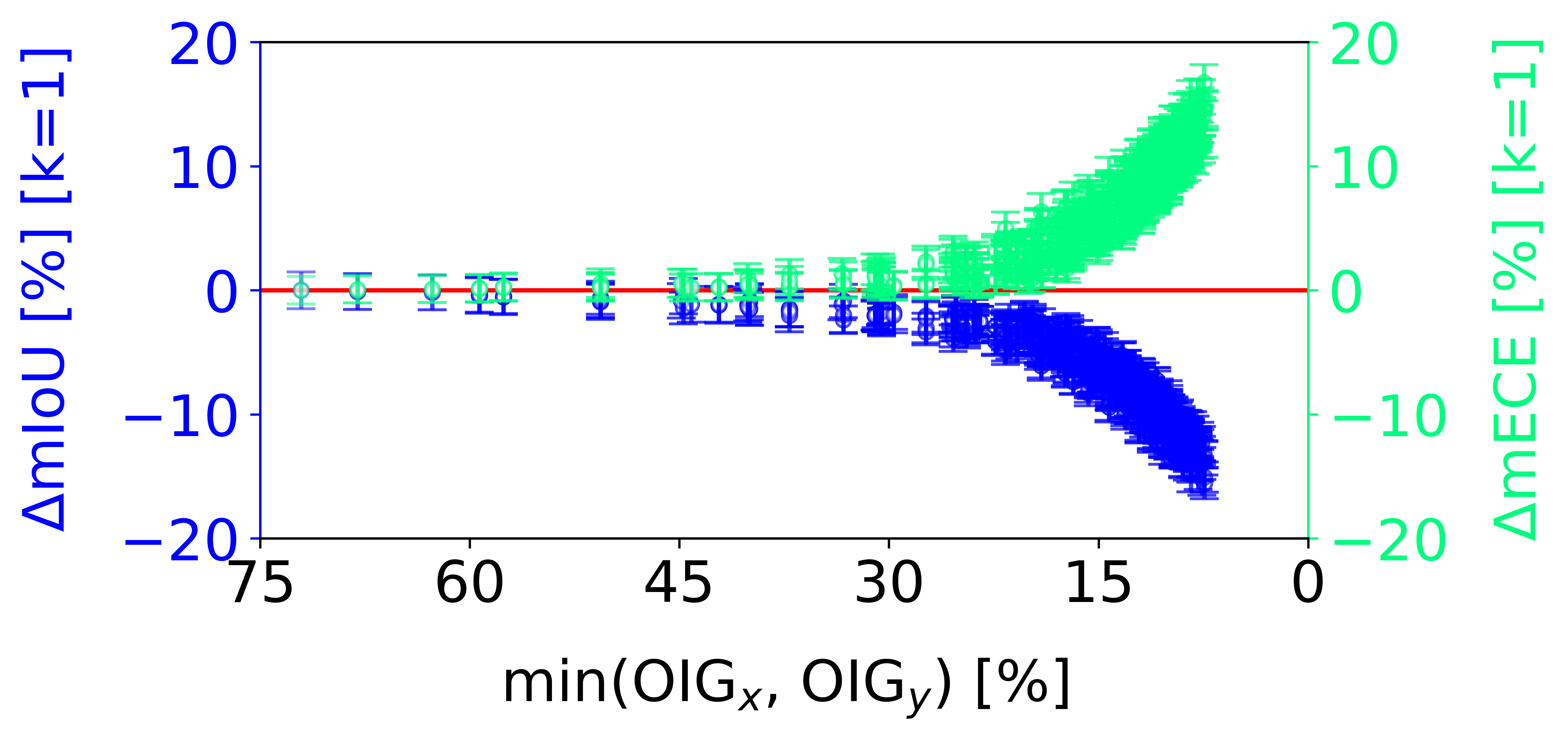}
      \hfill
      \includegraphics[width=0.\textwidth]{mIoU_vs_OIG.png}
\captionof{figure}{The dependency of the mIoU and the mECE on different optical merit functions is shown. The results are almost symmetrical around the baseline if the trend of the mIoU is considered in relation to the mECE, which is scientifically justified by Figure~\ref{fig:pearson_correlation_conf_vs_acc}. In summary, the refractive power and the MTF at half-Nyquist frequency do not demonstrate a functional relationship w.r.t.\ the mIoU and the mECE. In contrast, the Strehl ratio and the OIG indicate a functional relationship, which might even fulfill the required bijectivity criterion.}
\label{fig:Correlation_analysis_HRNet}
\end{center}
\end{strip}


\clearpage
\newpage
\bibliographystyle{ieee_fullname}
\bibliography{egbib}
\end{document}